\documentclass[aps,prb,twocolumn,superscriptaddress,noeprint]{revtex4-2}
\usepackage[utf8]{inputenc}
\usepackage[T1]{fontenc}
\usepackage{amsmath}
\usepackage{amsfonts}
\usepackage{amssymb}
\usepackage{graphicx}
\usepackage{dcolumn}
\usepackage{subcaption} 
\usepackage{comment}
\usepackage{cancel}
\usepackage{bm}
\usepackage{multirow}
\usepackage{mathdots}
\usepackage{amssymb}
\usepackage{hyperref}
\usepackage{xcolor}
\usepackage{soul}
\usepackage{physics}
\usepackage{MnSymbol}
\usepackage{float}
\usepackage{mathtools}
\usepackage[bb=boondox]{mathalfa}
\usepackage{dsfont}
\usepackage{comment}
\usepackage{slashed}
\usepackage{tikz}
\usepackage{bigints}
\usepackage{cleveref} 
\crefname{figure}{Fig.}{Fig.}
\crefname{equation}{}{}
\usepackage[normalem]{ulem}
\usepackage{enumitem}
\usepackage{array}

\newcommand{\lp}{\left(}
\newcommand{\rp}{\right)}

\begin{document}

\title{
Comparing quantum complexity and quantum fidelity}
\author{Nadir Samos S{\'a}enz de Buruaga}
\thanks{nadir.samos@tecnico.ulisboa.pt}
\affiliation{CeFEMA, LaPMET, Instituto Superior Técnico,
Universidade de Lisboa, Av. Rovisco Pais, 1049-001 Lisboa, Portugal}

\date{\today}

\begin{abstract}
Quantum complexity measures the difficulty of obtaining a given state starting from a typically unentangled state. In this work, we show that complexity, when defined through the minimization of a Riemannian cost functional over the manifold of Gaussian states, provides the same information as quantum fidelity and is therefore capable of detecting quantum phase transitions. However, it does not offer a more refined analysis than entanglement entropy for a given state. We conclude that incorporating a notion of spatial locality into the computation of complexity is essential to uncover new physics beyond what is accessible through entanglement entropy and fidelity.

\end{abstract}
\maketitle

\section{Introduction}
Consider an isolated quantum system of $N$ constituent qudits of local dimension $\mathbb{C}^d$. All states that may describe this system belong to the Hilbert space $\mathcal{H}$. 
As shown in Ref. \cite{poulinQuantumSimulationTimeDependent2011}, the set of \emph{physical} states includes all those that can be obtained through the unitary evolution of a reasonable Hamiltonian—with some notion of spatial locality—in polynomial time relative to the system size. This set occupies an exponentially small volume in $\mathcal{H}$. This fact raises the appealing question of how to find tools that allow one to determine whether a given state drawn from $\mathcal{H}$ is physical or not \cite{linRealImaginaryTimeEvolution2021}.

Quantum entanglement, the primary form of quantum correlations in pure states, sheds light on this problem. In terms of entanglement entropy (EE), we distinguish \emph{area-law} states \cite{eisertColloquiumAreaLaws2010} as those in which no connected subsystem $A$ has an EE that scales with its volume, $\text{Vol}(A)$. Although far from typical, this small subset is of crucial relevance, as it contains the ground and low-energy states of many physical Hamiltonians \cite{hastingsSolvingGappedHamiltonians2006}. However, the vast majority of quantum states conform to \emph{volume-law}. In fact, a randomly sampled state will likely have nearly maximal EE, following the so-called Page law \cite{pageAverageEntropySubsystem1993}, so it is not possible to determine whether a given state is physical or non-physical solely based on EE \cite{susskindEntanglementNotEnough2016}.

Quantum \emph{circuit complexity} is a strong candidate for elucidating this important problem. Drawing from classical computer science \cite{liIntroductionKolmogorovComplexity2008}, it addresses the challenge of building (within certain tolerance) a specific target unitary $U$ using a given set of $n$ universal quantum gates $\{u_i\}_{i=1}^n$. The \emph{state complexity} is therefore the unitary complexity of the least complex unitary that connects a target state from a typically unentangled reference state. 

The concept of complexity emerged within high-energy physics and holography, highlighting a contradiction in the AdS/CFT correspondence. While the wormhole volume increases exponentially over time, boundary-field theory observables, such as EE, reach saturation in polynomial time. To resolve this paradox, it was suggested that the dual of volume is the complexity $\mathcal{C}$ \cite{susskindComputationalComplexityBlack2016,brownHolographicComplexityEquals2016}. It is important to note that due to the vastness of the unitary group in a many-body quantum system, it is very unlikely that the unitaries sampled at time $T$ exactly correspond to the inverses of those applied at early times. Hence, it is highly improbable that the unitaries acting over time will completely cancel out the effects of the previous ones. Indeed, based on this intuition, a linear growth of complexity was conjectured \cite{brownSecondLawQuantum2018} and has recently been confirmed \cite{haferkampLinearGrowthQuantum2022,brandaoModelsQuantumComplexity2021}.

In addition to black hole physics, complexity has proven to be a versatile and profound concept that can illuminate the quantification of thermalization, randomness, and quantum chaos\cite{robertsChaosComplexityDesign2017,cotlerChaosComplexityRandom2017,hunter-jonesChaosRandomMatrices2018,aliChaosComplexityQuantum2020,brandaoModelsQuantumComplexity2021,nandyQuantumStateComplexity2024,oszmaniecSaturationRecurrenceQuantum2024,guoComplexityNotEnough2024}. Furthermore, it has been explored in the field of theoretical condensed matter as a tool to detect quantum phase transitions\cite{huangQuantumCircuitComplexity2015,liuCircuitComplexityTopological2019,aliPostquenchEvolutionComplexity2020,xiongNonanalyticityCircuitComplexity2020,caputaQuantumComplexityTopological2022,soodCircuitComplexityCritical2022,roca-jeratCircuitComplexityPhase2023,jayaramaRealspaceCircuitComplexity2023,palComplexityLipkinMeshkovGlickModel2023}.

Although the purpose of complexity is clear, there is no universally accepted definition of it. Circuit complexity has a well-defined operational significance; however, it is influenced by the choice of the set of native gates $\{u_i\}_{i=1}^{n}$ and the tolerance level \cite{brownComplexityGeometrySingle2019}. Another promising definition is the recently introduced \emph{spread complexity} \cite{balasubramanianQuantumChaosComplexity2022,caputaSpreadComplexityTopological2022,caputaQuantumComplexityTopological2022} which quantifies the complexity in terms of the size of the Krilov basis.

In this paper, we focus on the \emph{geometry of complexity} \cite{nielsenQuantumComputationGeometry2006,nielsenOptimalControlGeometry2006}, a geometric interpretation of circuit complexity that corresponds to the geodesic distance in a non-isotropic metric connecting the identity $\mathbb{1}$ to the desired unitary $U$. Crucially, the non-isotropic nature arises from imposed constraints, such as $k$-locality, which penalize certain directions in the metric. This approach to complexity has been explored in the context of quantum field theory \cite{jeffersonCircuitComplexityQuantum2017,khanCircuitComplexityFermionic2018,guoCircuitComplexityCoherent2018,bhattacharyyaCircuitComplexityInteracting2018,hacklCircuitComplexityFree2018,camargoComplexityNovelProbe2019,chapmanComplexityEntanglementThermofield2019,jiangCircuitComplexityFree2020} and conformal field theory \cite{camargoPathIntegralOptimization2019,floryGeometryComplexityConformal2020,bracciaComplexityPresenceBoundary2020}, and has also been extended to the realm of mixed states \cite{digiulioComplexityMixedGaussian2020}. In particular, we shall employ the covariance matrix approach developed in Ref. \cite{hacklCircuitComplexityFree2018}.

The quantum fidelity, which measures the closeness between quantum states, also provides a notion of distance in the Hilbert space. It is a fundamental concept in quantum information that has proven to be highly useful in various fields, including the detection of quantum phase transitions\cite{zanardiInformationTheoreticDifferentialGeometry2007,garneroneFidelityTopologicalQuantum2009,guFidelityApproachQuantum2012}, identifying signatures of quantum chaos\cite{gorinDynamicsLoschmidtEchoes2006,goussevLoschmidtEcho2012,yanInformationScramblingLoschmidt2020}, dynamical phase transitions\cite{heylDynamicalQuantumPhase2018}, and benchmarking quantum computers\cite{samosFidelityDecayError2024}, among others. 

However, both fidelity and complexity \emph{should} provide distinct notions of proximity or similarity \cite{brownComplexityGeometrySingle2019}. While fidelity assigns a maximal separation between two states describing a many-body quantum system $\ket{\Phi}$ combined with a spin-$1/2$ particle, $\ket{\tilde{\pm}} \equiv \ket{\Phi} \otimes \ket{\pm}$ complexity is expected to be more refined and to assign a smaller distance in this scenario.

The objective of this paper is to emphasize the importance of incorporating locality into the complexity metric to uncover new physics beyond quantum fidelity. The idea of adding penalty factors to prohibit $k-$ local interaction terms with $k>2$ was already considered in the seminal work \cite{nielsenQuantumComputationGeometry2006}. Here, we show that even when restricted to the realm of Gaussian states—a limited class of states generated by quadratic Hamiltonians—complexity yields insights analogous to those provided by fidelity. To illustrate this, we analyze two paradigmatic models in condensed matter physics: the Su–Schrieffer–Heeger (SSH) and Kitaev chains. Furthermore, we highlight how this notion of complexity remains unable to surpass the insights provided by the entanglement entropy.

The paper is organized as follows. In Sec. \ref{sec:setting_the_scene} we will formally introduce the complexity and fidelity in the context of Gaussian states, along with the two many-body models considered in this work. In Sec.\ref{sec:compvsfid} we compare both quantities and discuss their equivalence. In Sec.\ref{sec:compvsent}, we provide an example where complexity yields trivial information. We conclude this paper with a discussion of the main results presented and future work in Sec. \ref{sec:conclusions}.

\section{Setting the scene}
\label{sec:setting_the_scene}

In this work, we focus on the restricted subspace of Gaussian states. These states play a fundamental role in quantum information and condensed matter physics as they serve as simple models within the second quantization formalism, allow for analytical treatment, and form the basis of essential numerical mean-field methods. They are termed Gaussian states because they are fully characterized by the first and second correlation functions, with all higher-order correlations determined through Wick's theorem \cite{wickEvaluationCollisionMatrix1950}. These states correspond to the eigenvectors of so-called free Hamiltonians and can describe both bosonic and fermionic many-body systems. In this study, we concentrate on free fermionic systems; however, the conclusions drawn here should also be applicable to the bosonic framework.

Given a system of $N$ fermions satisfying the canonical anticommutation relations $\{c_i,c^\dagger_j\}=\delta_{ij}$, we may decompose each of them into Majorana fermions $c_i=1/2(\alpha_i+i\beta_i)$ that satisfy the anticommutation relations $\{\alpha_i,\beta_j\}=0$, $\{\alpha_i,\alpha_j\}=\{\beta_i,\beta_j\}=2\delta_{ij}$, and write the spinor
\begin{equation}
    \bm{\xi}=\begin{pmatrix}
        \alpha_1\\
        \vdots\\
        \alpha_N\\
        \beta_1\\
        \vdots\\
       \beta_N
    \end{pmatrix}= \begin{pmatrix}
        \bm{\alpha}\\
        \bm{\beta}
    \end{pmatrix},
    \label{eq:majo_spinor}
\end{equation}

The two-point correlation function, the primary quantity for these states, can be decomposed into symmetric and antisymmetric components \cite{hacklCircuitComplexityFree2018}:

\begin{equation}
    \ev{\xi_i\xi_j}=\frac{1}{2}\ev{\{\xi_i,\xi_j\}}+\frac{1}{2}\ev{[\xi_i,\xi_j]}.
    \label{eq:2point}
\end{equation}
In the case of fermions, the symmetric part is entirely determined by their anticommutation nature, with the non-trivial information encoded in the antisymmetric component. We refer to this as \emph{covariance matrix} $\bm{\Gamma}$.
\begin{equation}
    \Gamma_{ij}:=\frac{1}{2}\ev{[\xi_i,\xi_j]}
    \label{eq:defcovariance}
\end{equation}
In terms of the spinor \eqref{eq:majo_spinor}:
\begin{equation}
    \ev{\bm{\xi}\bm{\xi}^T}=\lp\begin{array}{c|c}
  \mathbb{1}& \mathbb{0} \\ 
  \hline
  \mathbb{0} &\mathbb{1}
 \end{array}\rp +i\bm{\Gamma}.
\end{equation}
Observe that $\bm{\Gamma}$ is real and antisymmetric, and its simplest form is obtained when considering the Fock vacuum (see Appendix \ref{app:oncov_andmodels} for more details):

 \begin{equation}
     \bm{\Gamma}_\text{Fock}=\lp\begin{array}{c|c}
  \mathbb{0}& \mathbb{1}\\ 
  \hline
  -\mathbb{1}& \mathbb{0}
 \end{array}\rp \ ,
 \label{eq:fock_cov}
 \end{equation}
 which corresponds to the standard form of symplectic matrices. At this point, we can make two observations. The first is that  ${\Gamma}^2_\text{Fock}=-\mathbb{1}$. The second is that every covariance matrix can be brought to the standard form \eqref{eq:fock_cov}  by a suitable Bogoliubov rotation $Q \in O(2N)$, i.e., a Gaussian transformation that preserves the anticommutation relations. Consequently, all pure fermionic states satisfy $\bm{\Gamma}^2 = -\mathbb{1}$ \cite{peschelReducedDensityMatrix2004}. It also follows that  
\begin{equation}
\bm{\Gamma}^{-1}=\bm{\Gamma}^T.
\end{equation}

After having defined the states, we can now proceed to present fidelity and complexity as instruments for assessing their closeness.
 
\subsection{Fidelity in Gaussian States}
\label{subsec:fidelity}
Quantum fidelity quantifies the proximity between quantum states. Let us consider two pure states, which we will refer to as \emph{reference state} $\ket{\Psi_R}$ and \emph{target state $\ket{\Psi_T}$} for future convenience. 
The quantum fidelity is defined as:
\begin{equation}
    \mathcal{F}(\xi_R,\xi_T)=|\braket{\xi_R}{\xi_T}|^2.
    \label{eq:fid}
\end{equation}

Fidelity also provides a notion of distance in the Hilbert space, known as the Fubini-Study distance 
\begin{equation}
    d_{FS}:= \arccos{\sqrt{\mathcal{F}}},
    \label{eq:fubini}
\end{equation}  
which corresponds to  the real part of the complex metric associated with the manifold of rays on $H$—that is, in the projective space $P\mathcal{H}$. It is also referred to as the Geometric Quantum Tensor \cite{provostRiemannianStructureManifolds1980,hetenyiFluctuationsUncertaintyRelations2023}.  

When we restrict ourselves to states that depend on a finite set of parameters, it is known that it is possible to detect quantum phase transitions by finding singularities in the (pulled-back) metric on the manifold of parameters \cite{zanardiInformationTheoreticDifferentialGeometry2007,guFidelityApproachQuantum2012}. The fidelity between Gaussian states has been computed by several authors, with existing expressions in terms of the BCS formulation of the many-body state \cite{robledoPracticalFormulationExtended1994, mbengQuantumIsingChain2024}, as well as in terms of the covariance matrix \cite{banchiQuantumFidelityArbitrary2015,swingleRecoveryMapFermionic2019}.

In particular, we will use the expression for two fermionic Gaussian states \cite{bravyiLagrangianRepresentationFermionic2005}:
\begin{equation}
     \mathcal{F}(\xi_R,\xi_T)=\sigma\text{Pf}\lp\frac{\bm{\Gamma}_R+\bm{\Gamma}_T}{2}\rp,
\end{equation}
where $\text{Pf}(A)$ refers to the Pfaffian of the skew-symmetric matrix $A$, and $\sigma=\pm1$ refers to the parity of both states. Indeed, if the states have different parities, the Pfaffian vanishes since the largest minor is odd-dimensional.

\subsection{Complexity in Gaussian States}
\label{subsec:complexity}
As mentioned in the introduction, circuit complexity measures the minimum number of gates required to approximate a given unitary within a certain tolerance. In the realm of Gaussian states, one has access to matchgates—classically simulable two-qubit gates that act on nearest-neighbor qubits. These gates are characterized by a set of linear algebraic constraints related to graph matchings in combinatorial theory \cite{valiantQuantumCircuitsThat2012,terhalClassicalSimulationNoninteractingfermion2002}. By leveraging these properties, one can attempt to determine the optimal gate composition for a given unitary transformation. However, similar to the situation with general circuits, once a circuit is identified, demonstrating its optimality poses a challenge. As a result, there has been a growing interest in the continuum approach.

Concerning the geometrized formulation of complexity proposed in Ref. \cite{nielsenQuantumComputationGeometry2006}, the elementary universal gates are replaced by a continuous description, $u_i = e^{-i\epsilon h_i}$, where $\{h_i\}_{i=1}^n$ forms a basis of Hermitian operators, and $\epsilon \to 0$. In this framework, an intermediate unitary $U(s)$ is described as the continuous evolution of a time-dependent Hamiltonian, with ordering from right to left in its construction:  

\begin{equation}
    U(t) = \overleftarrow{\mathcal{T}} e^{-i\int_0^t d\tau H(\tau)},
\end{equation}
with the constraints $U(t=0) = \mathbb{1}$ and $U(t=1) = U_T$. The time-dependent Hamiltonian is expressed as  $H(\tau) = \sum_{i} \alpha_i(\tau) h_i$. The set of coefficients $\alpha_i(\tau)$ at a given time $\tau$ defines the tangent vectors on the manifold, denoted as $\boldsymbol{\alpha}(\tau) = (\alpha_1(\tau), \dots, \alpha_N(\tau))$.

The geometry of complexity is obtained by minimizing an appropriate length functional $F[U(\tau),\boldsymbol{\alpha}(t)]$
\begin{equation}
    C_F[U_T] = \min_{\{U(\tau)\}} \int dt \, F[U(t), \boldsymbol{\alpha}(t)].
\end{equation}

If $F[U(\tau),\boldsymbol{\alpha}(t)]$ defines a Finsler metric—a generalization of the Riemannian metric (see Ref. \cite{chapmanQuantumComputationalComplexity2022} and the discussion therein about different choices of $F[U(\tau),\boldsymbol{\alpha}(t)]$)—then the problem of finding the complexity then reduces to identifying the geodesics of this metric. The significant insight provided in \cite{hacklCircuitComplexityFree2018} lies in further imposing that the Hermitian operators $\{h_i\}$ form the Lie algebra of the group $O(2N)$ corresponding to the Bogoliubov transformations of $N$ fermions. Determining the state complexity requires the additional optimization of finding the least complex unitary that connects $\ket{\psi_T}$ to $\ket{\psi_R}$.
\begin{equation}
    \mathcal{C}(\Psi_R,\Psi_T)=\min_{\ket{\Psi_T}=U\ket{\Psi_R}} \mathcal{C}(U)
    \label{eq:def_statecomplexity}
\end{equation}
In the group theoretic approach, this is achieved by finding the geodesics not in $O(N)$ but in the quotient manifold with the stabilizer group that contain all the unitaries that leave the state invariant $O(N)/U(N)$. Furthermore, when the length functional corresponds to the usual Riemmanian distance
\begin{equation}
F[U(t), \boldsymbol{\alpha}(t)] \equiv \sqrt{\sum_i \alpha_i^2},
\label{eq:riemmanian_cost}
\end{equation}
the complexity between two Gaussian states $\xi_R$ and $\xi_T$ is given by
\begin{equation}
    \mathcal{C}(\xi_R,\xi_T)=\norm{\frac{1}{2}\log\bm{\Delta}}=\frac{1}{2}\sqrt{\Tr(i\log\bm{\Delta})^2},
    \label{eq:complexity_cov}
\end{equation}
where $\bm{\Delta}$ is called the relative covariance matrix
\begin{equation}
    \bm{\Delta}\equiv \bm{\Gamma_T}\bm{\Gamma_R}^{-1}=\bm{\Gamma_T}\bm{\Gamma_R}^T=-\bm{\Gamma_T}\bm{\Gamma_R},
    \label{eq:rel_cov}
\end{equation}
that contains all the invariant information between target and reference states.

\subsection{Models}
\label{sec:models}
On this work we consider two prototypical free-fermion 1D models: The Su–Schrieffer–Heeger (SSH) and Kitaev chains. Both show a trivial and a nontrivial phase with a critical point described by a conformal field theory of central charge $c=1$ and $c=1/2$, respectively. 
\subsubsection{SSH Hamiltonian}
The SSH \cite{suSolitonsPolyacetylene1979} model is the most prominent example of a topological insulator and was originally proposed to describe the physics of polyacetilene. It describes a system of $N$ complex fermions given by the Hamiltonian
\begin{equation}
    H_{\text{SSH}}=-\sum_{m=1}^{N-1}(1+(-1)^m\delta)\left(c^\dagger_mc_{m+1} \ + \ \text{h.c}\right),
    \label{eq:ssh_ham}
\end{equation}
for the dimerization parameter $\delta\in[-1,1]$, where $c_i$ ($c^\dagger_i$) are annihilation (creation) operators satisfying the canonical anticommutation rules $\{c_i,c^\dagger_j\}=\delta_{ij}$. The model is dual to an XY dimerized spin $1/2$ chain via a Jordan-Wigner transformation.
The model presents a sublattice (or chiral) symmetry, as well as particle and time reversal symmetry, thus belonging to the BDI class of the Altland-Zirnbauer classification \cite{altlandNonstandardSymmetryClasses1997}. The case $\delta=0$ corresponds to the translational-invariant model whose low-energy continuum limit is described by a conformal field theory of central charge $c=1$.

\subsubsection{Kitaev Hamiltonian}
The Kitaev model corresponds to a toy model of a one-dimensional p-wave superconductor \cite{kitaevUnpairedMajoranaFermions2001}. It describes $N$ complex fermions given by the Hamiltonian
\begin{equation}
   H_\text{K}= -\sum_{m=1}^{N-1}J\left(c^\dagger_mc_{m+1} + c^\dagger_mc^\dagger_{m+1} \ + \ \text{h.c}\right) + 2\sum_{m=1}^{N}hc^\dagger_mc_{m},
   \label{eq:kitaev_ham}
\end{equation}
with $J=1/2(1+\delta)$ and $h=1/2(1-\delta)$.
The model is dual to the spin $1/2$ quantum transverse-field Ising model via a Jordan-Wigner transformation, but it takes its simplest form when it is written in terms of Majorana fermions. In this basis, the above system is described as a Majorana chain with first neighbors alternating couplings:
\begin{equation}
    H_\text{K}=\frac{i}{4}\sum_{m=1}^{N-1}(1-\delta)\alpha_m\beta_m+ (1+\delta)\beta_m\alpha_{m+1} +(1-\delta)\alpha_N\beta_N,
\end{equation}
which attains the same functional form of the SSH model \eqref{eq:ssh_ham}. 

\begin{figure}
    \centering
\includegraphics[width=1\linewidth]{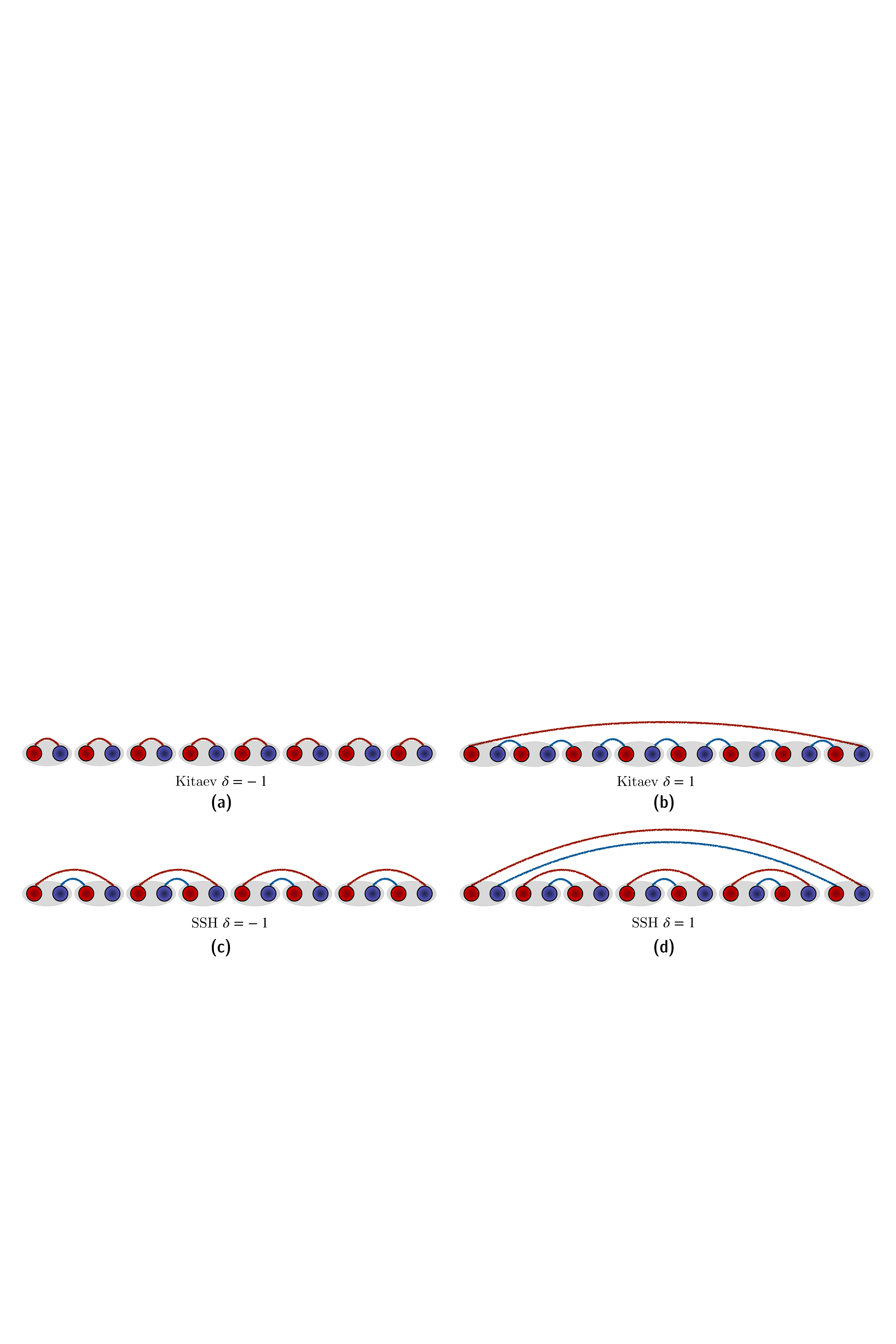}
    \caption{Ground states of the extremal cases $\delta=\pm1$ of the SSH \eqref{eq:ssh_ham} and Kitaev \eqref{eq:kitaev_ham} chains of $8$ sites in terms of Majorana fermions (see Appendix \ref{app:oncov_andmodels} for more details) where the complex fermion $c_i=1/2(\alpha_i+i\beta_i)$ corresponds to the shaded grey area. The EE of cutting each bond is $\log(d_q)$ where $d_q=\sqrt{2}$ is the quantum dimension of the Majorana fermion \cite{bonesteelInfiniteRandomnessFixedPoints2007}.}
    \label{fig:ssh_kitaev}
\end{figure}

The ground states of both models corresponding to the cases $\delta=\pm=1$ can be seen represented in Fig. \ref{fig:ssh_kitaev}. Since we chose the Majorana basis as the common one to write covariance matrices, we shall rewrite the SSH ground states in this new basis. The valence bond state describing the cases $\delta=\pm1$ can be written in terms of generalized bonds \cite{bonesteelInfiniteRandomnessFixedPoints2007}. We refer to Appendix \ref{app:oncov_andmodels} for more details. 

\section{Complexity vs Fidelity}
\label{sec:compvsfid}
To make a fair comparison between both quantities, we start by writing the fidelity \eqref{eq:fid} in terms of the covariance matrix \eqref{eq:rel_cov}. It is easy to see that
\begin{equation}
\begin{split}
\mathcal{F}(\xi_R,\xi_T)=\sigma\text{Pf}\lp\frac{\bm{\Gamma}_R+\bm{\Gamma}_T}{2}\rp=\sigma\sqrt{\det\lp\frac{\bm{\Gamma}_R+\bm{\Gamma}_T}{2}\rp}\\
=\sigma\sqrt{\det{\frac{\mathbb{1}+\bm{\Delta}}{2}}\bm{\Gamma_R}} = \sigma\det{\sqrt{\frac{\mathbb{1}+\bm{\Delta}}{2}}},
\end{split}
\label{eq:fid_relcov}
\end{equation}
where we have used that $\det\bm{\Gamma}=1$.

Returning to the definition of $\bm{\Delta}$ \eqref{eq:rel_cov} and the properties of the covariance matrix, it is clear that $\bm{\Delta}\bm{\Delta}^T=\bm{\Delta}^T\bm{\Delta}=\mathbb{1}$ and consequently $\bm{\Delta}\in O(2N)$. According to the spectral theorem, the spectrum of $\bm{\Delta}$ must lie in the unit circle, with complex roots that come in conjugate pairs: $\{\pm1,e^{i\theta},e^{-i\theta}\}$ with $\theta\in(0,\pi)$. Moreover, observe also that using \eqref{eq:fock_cov},
\begin{equation}
\bm{\Gamma}_{\text{Fock}}\bm{\Delta}\bm{\Gamma}^T_{\text{Fock}}=\bm{\Delta}^T=\bm{\Delta}^{-1}.
\end{equation}

\begin{figure}
    \centering
    \includegraphics[width=0.93\linewidth]{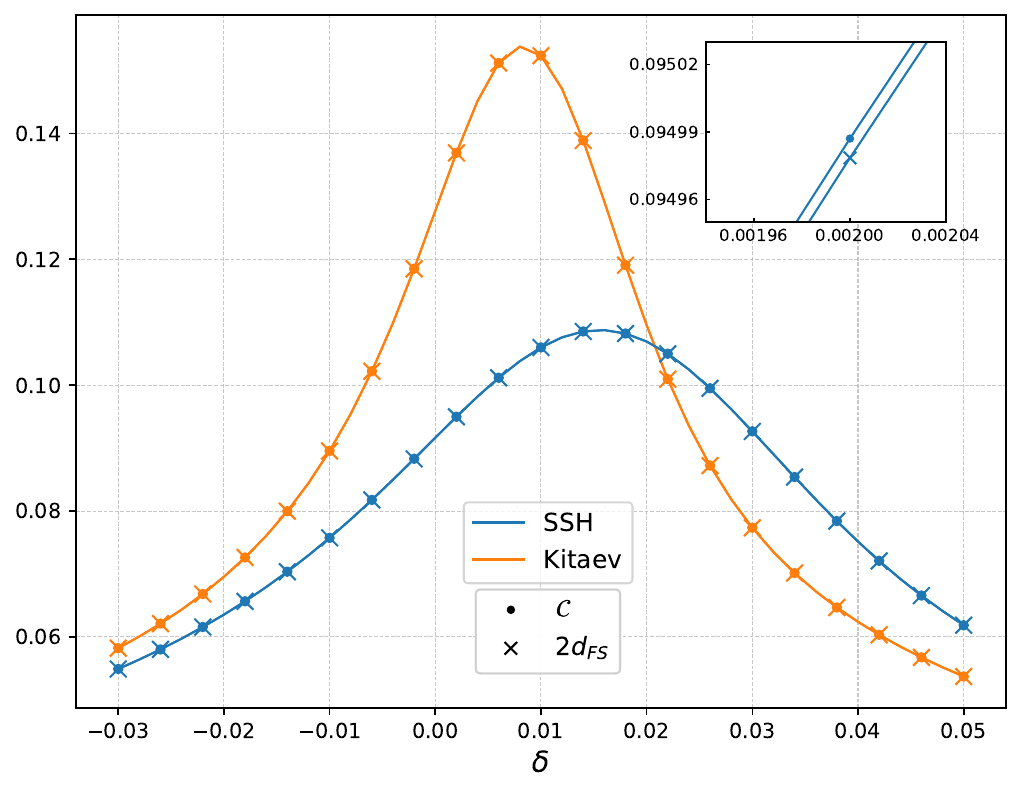}\\
    \includegraphics[width=0.95\linewidth]{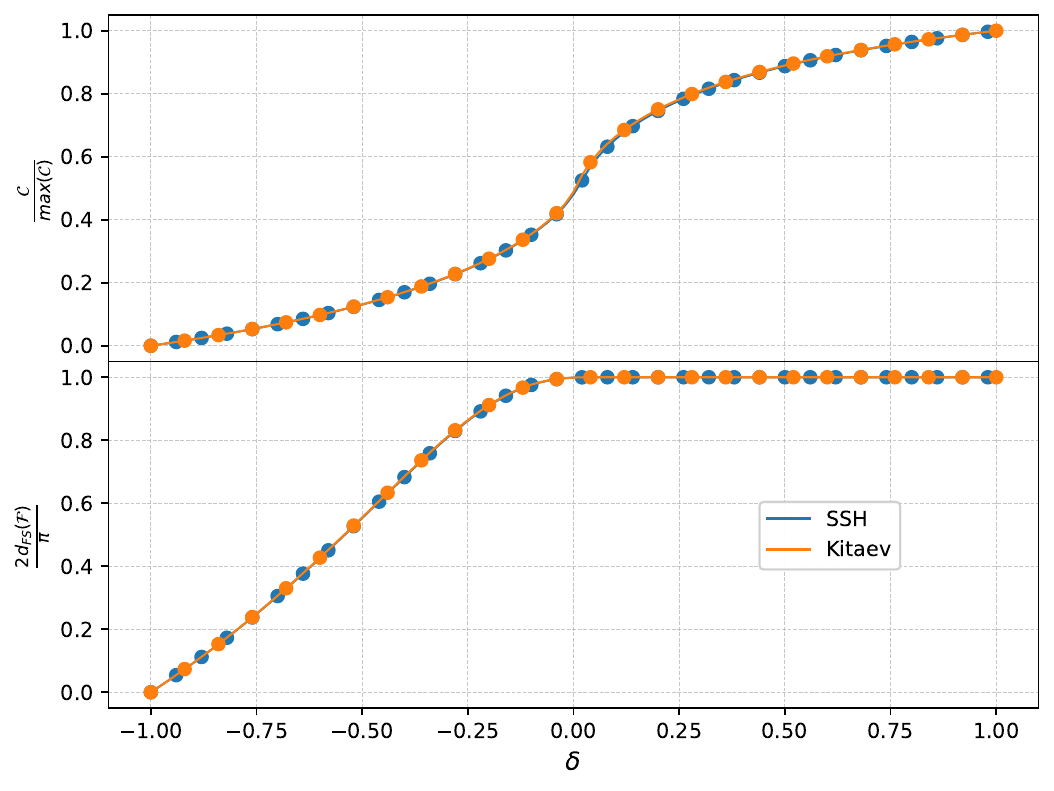}\\
    \includegraphics[width=0.95\linewidth]{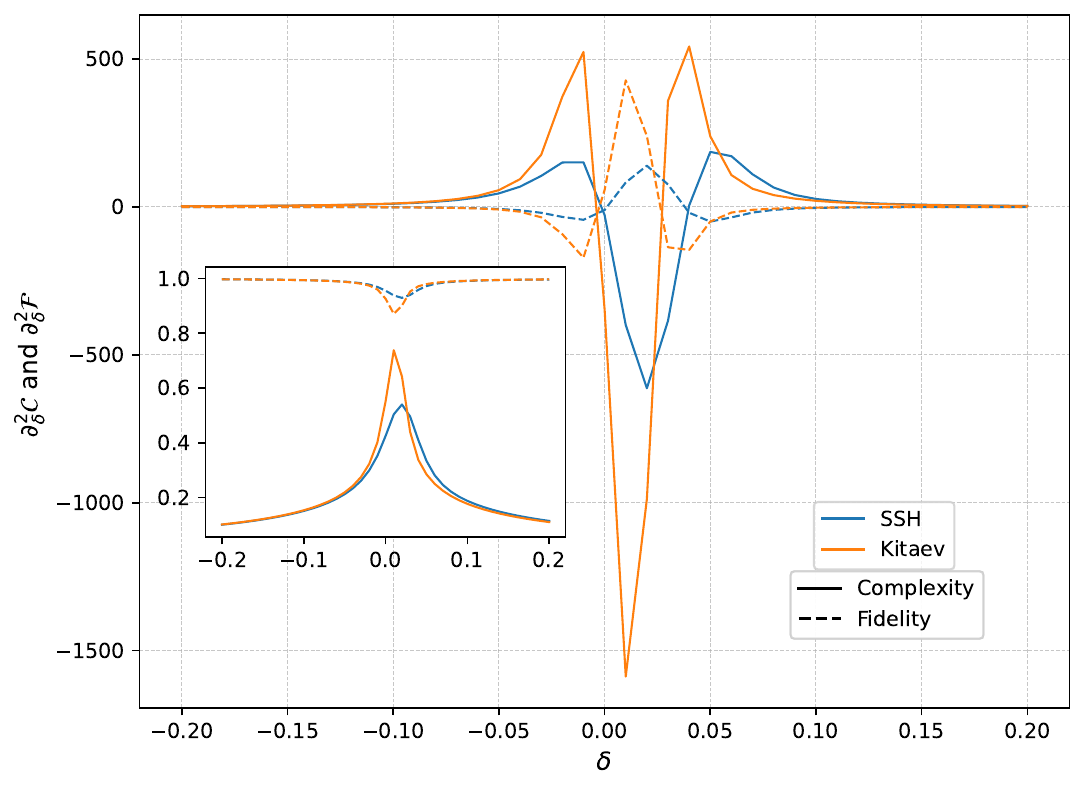}\\
    \caption{Quantum Complexity ($\mathcal{C}$) and Fubini-Study ($d_{FS}$) distances as a function of $\delta$ for both ground states of the SSH and Kitaev chains with $N=124$ Dirac fermions. Top: supporting evidence of expression \eqref{eq:relation_comp_dfs} for ground states of very close Hamiltonians ($\epsilon=0.002$).
    Middle: reference states are given by the trivial valence-bond states (see Fig. \ref{fig:ssh_kitaev} (a) and (c)), whereas $\xi_T$ varies with $\delta$. Observe that both measures do not distinguish between the two models. Bottom: second derivative of the fidelity and the complexity for similar ground states. Both quantities show a singularity around the critical point $\delta=0$ with finite size deviations \cite{camposvenutiQuantumCriticalScaling2007}.}
    \label{fig:comp_fid}
\end{figure}

Therefore, in analogy to the symplectic constraint, given an eigenvalue $\lambda$, it follows that $1/\lambda$ is also an eigenvalue. Hence, putting both conditions together, we find that the phases of the spectrum of the relative covariance matrix are given by tuples $\{0,0\}$, $\{\pi,\pi\}$ and quadruplets $\{\theta_j,\theta_j,-\theta_j,-\theta_j \}$ with $\theta_j\in (0,\pi)$. Hence, we can write the fidelity and the complexity in terms of these phases $\{\theta_j\}$.

From \eqref{eq:fid_relcov} we obtain
\begin{equation}
    \mathcal{F}(\xi_R,\xi_T)=\prod_{j=1}^{2N}\sqrt{\frac{1+e^{i\theta_j}}{2}}=\prod_{j=1}^{2N}e^{i\theta_j/4}\sqrt{\cos\frac{\theta_j}{2}}
\label{eq:fid_eigenvalues}
\end{equation}

Concerning the complexity, we find from \eqref{eq:complexity_cov} that
\begin{equation}
    \mathcal{C}(\xi_R,\xi_T)=\sqrt{\sum_{j=1}^{2N} \lp\frac{\theta_j}{2}\rp^2}
    \label{eq:comp_eigenvalues}
\end{equation}

As discussed in Ref. \cite{hacklCircuitComplexityFree2018}, a given $\bm{\Delta}$ whose spectrum contains an odd number of pairs $\{\pi,-\pi\}$— such that they cannot be grouped into quadruplets — indicates that the states cannot be connected by a Bogoliubov transformation, as they belong to different parity sectors.

Observe the similarities between the two quantities \eqref{eq:comp_eigenvalues} and \eqref{eq:fid_eigenvalues} when expressed in the same terms. By taking into account the structure of the spectrum, we can simplify both quantities
\begin{equation}
\begin{split}
    \mathcal{F}(\xi_R,\xi_T)=\prod_{I=1}^{M}\cos^2\lp\frac{\theta_j}{2}\rp\\
     \mathcal{C}(\xi_R,\xi_T)=2\sqrt{\sum_{j=1}^{M} \lp\frac{\theta_j}{2}\rp^2}
\end{split}
\label{eq:fid_and_comp}
\end{equation}
where $M$ is the total number of quadruplets. 

First we shall show that when all the phases $\{\theta_j\}$ are all small the complexity and the Fubini-Study distance are proportional:
\begin{equation}
    \mathcal{C}(\xi_R,\xi_T)\approx 2 d_{FS}(\xi_R,\xi_T)
    \label{eq:relation_comp_dfs}
\end{equation}

To see it we keep the leading order in the expansion in \eqref{eq:fubini}:
\begin{equation}
\begin{split}
    \cos d_{FS}\approx 1-\frac{1}{2}d^2_{FS}&= \prod_{j=1}^{M}\lp 1-\frac{\theta^2_j}{8}\rp\approx 1-\frac{1}{2}\sum_{j=1}^M\lp\frac{\theta_j}{2}\rp^2\\
&d_{FS}\sim\sqrt{\sum_{j=1}^M\lp\frac{\theta_j}{2}\rp^2}.
\end{split}
\end{equation}
The above conditions are satisfied when $\xi_R$ and $\xi_T$ are very similar. This is illustrated in the top panel of Fig. \ref{fig:comp_fid}, where we show that the expression \eqref{eq:relation_comp_dfs} holds for the Kitaev and SSH Hamiltonians with $N=124$ Dirac fermions, specifically when considering ground states of very similar Hamiltonians with parameters $\delta$ and $\delta + \epsilon$, where $\epsilon \ll 1$.

However, when the states are not close enough, the complexity typically scales with the size of the system, and the expression \eqref{eq:relation_comp_dfs} is not true in general. Indeed, the complexity can be more precise than the Fubini-Study distance, as can be seen in the middle panel of Fig. \ref{fig:comp_fid} where we consider in this case the reference states to be fixed to the clean representatives of the trivial phase ($\delta=-1$) that are illustrated in Fig. \ref{fig:ssh_kitaev} (a) and (c). Observe that the complexity follows the intuition: taking as reference the trivial state, the highest complexity is obtained when we aim to reach the non-trivial one. Interestingly, observe that both distances cannot distinguish between models as in the previous case (top panel).

Concerning the ability of complexity to detect phase transitions, it has been shown in Refs. \cite{liuCircuitComplexityTopological2019,xiongNonanalyticityCircuitComplexity2020} that the complexity is indeed capable of probing the phase transition of the Kitaev model. However, the case of the SSH model has been subject to some controversy regarding its detectability \cite{caputaSpreadComplexityTopological2022,aliTimeEvolutionComplexity2019}. This debate has motivated the exploration of an alternative definition of complexity, known as spread complexity \cite{caputaSpreadComplexityTopological2022} in the context of investigating quantum phase transitions. 

In addition to the evidence provided by the relation \eqref{eq:relation_comp_dfs}, we show here that the covariance matrix approach to complexity \cite{hacklCircuitComplexityFree2018} is as sensitive as fidelity in detecting phase transitions.
    
Without entering into the discussion of whether complexity is a universal probe of quantum phase transitions, here we shall assume that both the reference and target states are ground states of a Hamiltonian tunable by a parameter $\delta$, and that there exists a critical value $\delta_c$ at which the complexity shows non-analytic behavior: the derivative with respect to $\delta$ vanishes, $\dot{\mathcal{C}}(\delta_c)=0$, and the second derivative diverges, $\ddot{\mathcal{C}}(\delta_c)\to\infty$. As we show in Appendix~\ref{app:complexity_and_fidelity}, these conditions imply that $\dot{\theta_j}(\delta_c)=0$ for all $j$ and that at least one quadruplet diverges, $\theta_j(\delta_c)\to\infty$. However, when these conditions are satisfied, the fidelity also exhibits the same non-analyticity, with $\dot{\mathcal{F}}(\delta_c)=0$ and $\ddot{\mathcal{F}}(\delta_c)\to\infty$, meaning that all phase transitions detectable by (this definition of) complexity are also assessed by fidelity. In the bottom panel of Fig.~\ref{fig:comp_fid}, we present the second derivatives—also known as susceptibilities—of the fidelity the complexity. Note that, although they differ in magnitude, they exhibit divergent behavior with the same finite-size scaling deviations.

\begin{figure}
    \centering
    \includegraphics[width=0.90\linewidth]{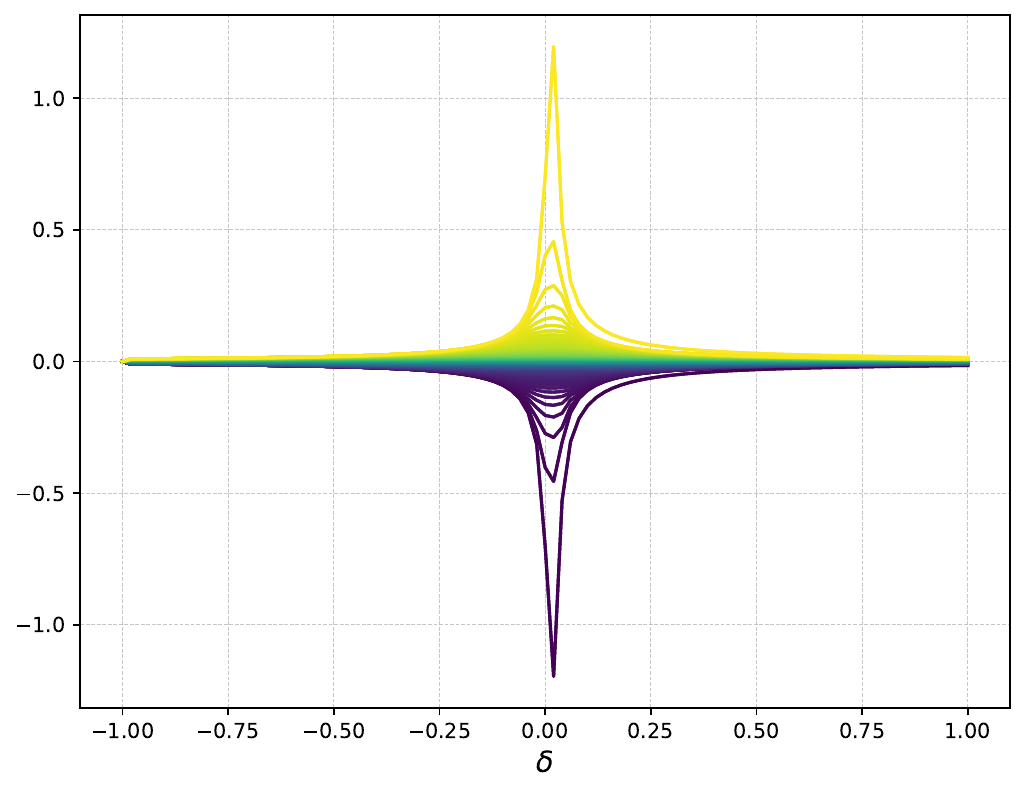}\\
        \includegraphics[width=0.90\linewidth]{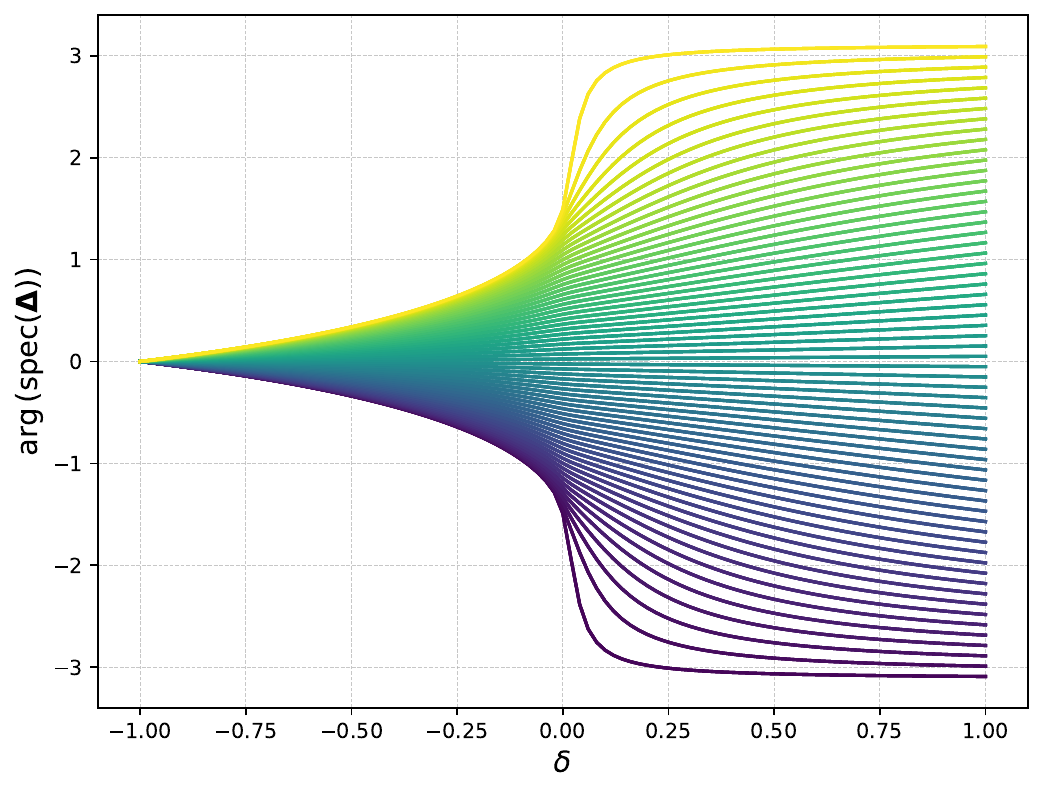}
    \caption{Phases $\{\theta_j\}$ of the spectrum of the relative covariance matrix $\bm{\Delta}$. Top panel illustrates the case when $\xi_R$($\xi_T$) is the ground state of $H_K(\delta)$($H_K(\delta+\epsilon)$). Bottom panel corresponds to the case where $\xi_R$ is fixed to be the ground state of $H_{\text{SSH}}(\delta=-1)$.}
    \label{fig:spectrums}
\end{figure}

Finally, we emphasize that, just as the entanglement entropy encodes partial information from the entanglement spectrum \cite{liEntanglementSpectrumGeneralization2008a}, both the complexity and the fidelity also encode partial information about the spectrum of the relative covariance matrix. In Fig. \ref{fig:spectrums} we illustrate the evolution of the phases $\{\theta_j\}$. In the top panel, we present the case in which the reference and target states are very similar ground states of the Kitaev Hamiltonian \eqref{eq:kitaev_ham}. In the bottom panel, we illustrate the case of the SSH, where the reference state is fixed to be the trivial state $\delta=-1$. Note that the particle-hole symmetry of the spectrum is evident in the symmetry of the eigenvalues around the real axis. Additionally, observe that near the critical point $\delta\sim0$, the spectrum exhibits nonanalyticity in at least one quadruplet.

\section{Complexity and entanglement}
\label{sec:compvsent}
As we have shown in the previous section, it is noteworthy that both the fidelity and the aforementioned definition of complexity do not distinguish between the two models. In fact, they are connected; broadly speaking, the ground state of the SSH model is a valence bond state of Dirac fermions, while the Kitaev chain represents a generalized valence bond state of Majorana fermions \cite{bonesteelInfiniteRandomnessFixedPoints2007, samosEntanglementNoncriticalInhomogeneous2021}. Although the entanglement structures of the two states differ, they are closely related. In the language of spin chains, it has been shown \cite{igloiExactRelationshipEntanglement2008} that the entanglement entropies of the XY and transverse-field Ising models are connected. Specifically, for a chain of $L$ spins and a subsystem of length $\ell$, the entanglement entropies satisfy the relation $S_{SSH}(2\ell,2L) = 2 S_{\text{Kitaev}}(\ell, L)$.
 However, as we emphasized in the Introduction, complexity is expected to provide information when entanglement, or more precisely, entanglement entropy, cannot \cite{susskindEntanglementNotEnough2016}.

Notice that the trivial state of the Kitaev model corresponds to the maximally occupied state $\prod c^\dagger_i\ket{0}$ (see Fig.\ref{fig:ssh_kitaev} (a)) which is a typical unentangled state often used as a reference. In this section, we will show that the complexity of states with a fixed number of particles depends exclusively on the number of particles of the state.

To proceed, we consider the covariance matrix of the Fock vacuum \eqref{eq:fock_cov} and the generic covariance matrix of a state with a fixed particle number and the same parity.
\begin{equation}
    \bm{\Gamma}_{pc}=\lp\begin{array}{c|c}
  \mathbb{0}& \mathbb{1}-2\bm{C} \\ 
  \hline
  -\mathbb{1}+2\bm{C} & \mathbb{0}
 \end{array}\rp.
\end{equation}
Then, the relative covariance takes the form 
\begin{equation}
    \bm{\Delta}_{pc}=\lp\begin{array}{c|c}
  \mathbb{1}-2\bm{C}& \mathbb{0}\\ 
  \hline
  \mathbb{0} & \mathbb{1}-2\bm{C}
 \end{array}\rp.
\end{equation}
Since $\bm{C}$ fully describes the state \cite{peschelCalculationReducedDensity2003}, it always has half of its eigenvalues equal to $0$ and the other half equal to $1$. Hence, the spectrum of $\bm{\Delta}_{pc}$ is $\pm 1$, with each degeneracy $N$. Thus, it follows that the complexity \eqref{eq:comp_eigenvalues}
\begin{equation}
    \mathcal{C}(\ket{0},\ket{\Psi_{pc}}) = \sqrt N \dfrac{\pi}{2}
    \label{eq:comp_particle_conserving}
\end{equation}
is insensitive to the entanglement structure of $\ket{\Psi}_{pc}$. We deduce from \eqref{eq:comp_particle_conserving} that the Fock vacuum (and the fully occupied state) occupy a privileged position in the manifold such that all possible states of $N$ particles are equidistant from it, regardless of their entanglement structure. This occurs because an optimal Gaussian transformation can always be found, even if it is highly non-local. In this sense, we cannot conclude that complexity provides more information than entanglement entropy.

To verify this, we examine the half-chain entanglement entropy of the ground states of various particle-conserving models and compare it with the complexity of creating these states from the reference Fock vacuum with the same parity. In addition to the SSH model discussed above, we shall also consider the rainbow Hamiltonian \cite{ramirezEntanglementRainbow2015,rodriguez-lagunaMoreRainbowChain2017}  that describes an inhomogeneous free fermion system with hoppings that decay exponentially from the center at a rate determined by the parameter $h$
\begin{equation}
\begin{split}
    H_{\text{rainbow}}&=-\sum_{m=-L+1}^{L-1}J_m(c^\dagger_{m-1/2}c_{m+1/2}+ \text{h.c.}),\\
    &\quad\text{with}\quad\quad \begin{cases}
        J_{0}=e^{-h/2}\\
        J_{m\neq0}=e^{-h|m|}.
    \end{cases}
    \end{split}
    \label{eq:rainbow_ham}
\end{equation}

For strong inhomogeneity $h>0$ , the ground state is a concentric valence bond state that resembles a rainbow. As a consequence, the EE is maximal in the half-chain. For smaller inhomogeneity, it can be seen that the low-energy physics is described by a CFT $c=1$  in curved space-time. In this case, the entanglement entropy is linear and proportional to the inhomogeneity \cite{rodriguez-lagunaMoreRainbowChain2017}.

In addition, we compare two random models. The first model is also a chain with nearest-neighbor couplings, again given by \eqref{eq:rainbow_ham}, but in this case the couplings $J_m \in [0,1]$ are sampled uniformly and independently. The second model is defined by
\begin{equation}
    H_\text{random} = \bm{c}^\dagger \bm{T} \bm{c} = \sum_{i,j} T_{ij} c_i^\dagger c_j,
    \label{eq:Hrandom}
\end{equation}
where $\bm{T} \in \mathrm{GUE}(N)$ is an $N \times N$ random matrix drawn from the Gaussian Unitary Ensemble, with an additional sublattice symmetry constraint. This constraint is imposed by requiring that, given $\tau_z = \mathbb{1}_{N/2} \otimes \sigma_z$, we have $\tau_z \bm{T} \tau_z = \bm{T}$. This condition is applied for convenience, ensuring that all sampled cases yield a ground state with the same filling.

\begin{figure}[h!]
    \centering
\includegraphics[width=0.92\linewidth]{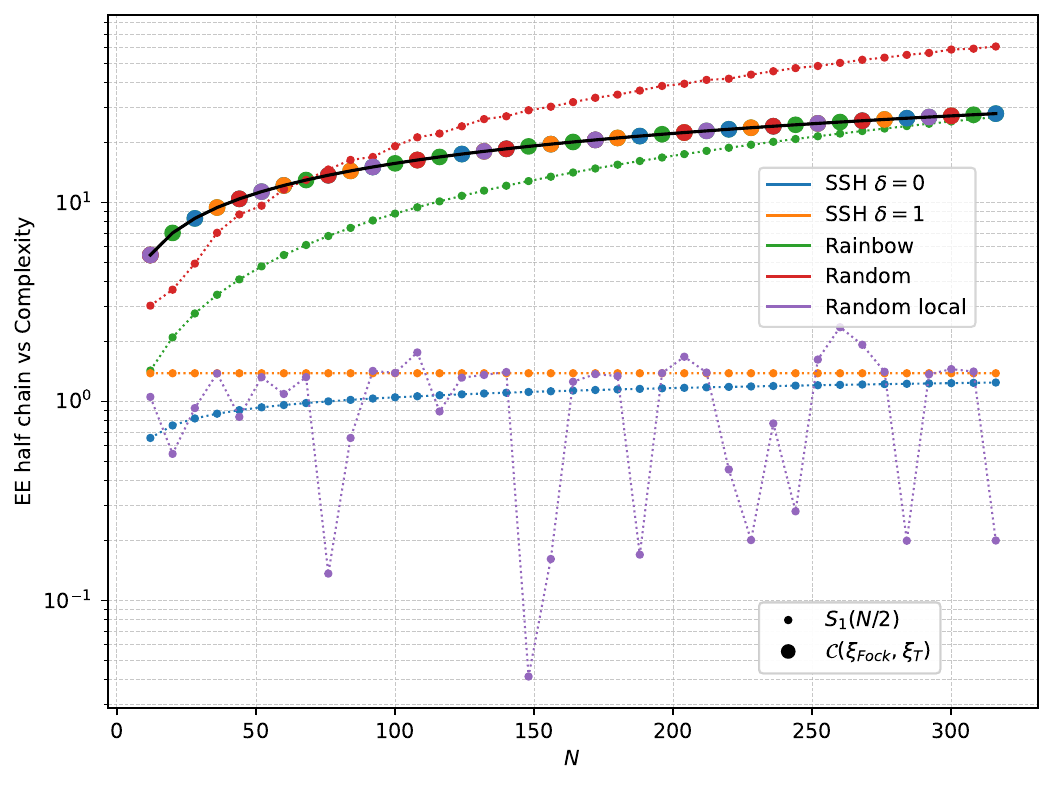}\\
\includegraphics[width=0.90\linewidth]{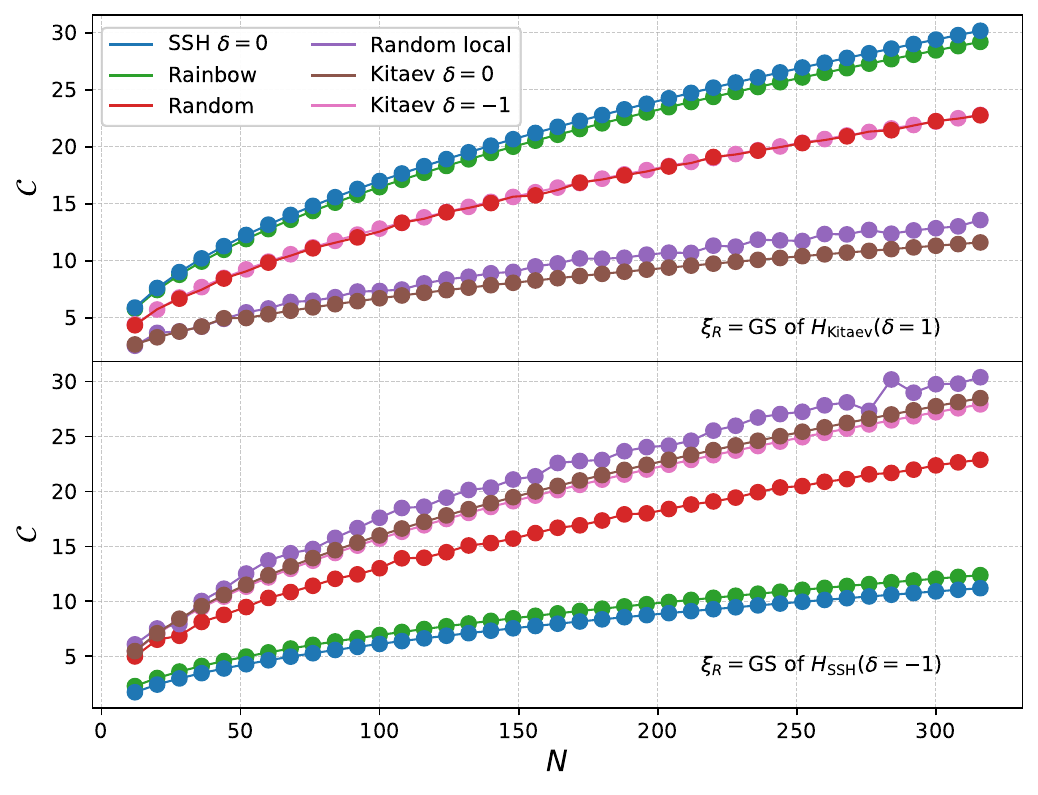}\
    \caption{Top: half-system entanglement entropy of different ground states: the SSH model \eqref{eq:ssh_ham} with $\delta=1$ and $\delta=0$, a random non-local Hamiltonian with chiral symmetry, a random-local Hamiltonian, and the rainbow state with inhomogeneity $h=1$. Complexity of reaching all these different states starting from the Fock vacuum. Black line is given by Eq. \eqref{eq:comp_particle_conserving}. Bottom: complexities of reaching the same ground states but starting from less privileged but still low-entangled reference states Fig. \ref{fig:ssh_kitaev}(b) and (c).}
    \label{fig:comp_vs_ee}
\end{figure}

In Fig. \ref{fig:comp_vs_ee} we plot the scaling of both the complexity and the EE for all the ground states. Observe that the complexity remains the same for all models and is given by \eqref{eq:comp_particle_conserving}.

It is also interesting to consider less-privileged reference states. In this case, obtaining an analytical prediction becomes more challenging, but we can rely on numerical results presented in the bottom panel of Fig.~\ref{fig:comp_vs_ee}. There, we show the complexities associated with reaching the ground states of the models discussed, starting from short-range entangled states such as the ground state of $H_{\text{K}}(\delta = 1)$, illustrated in Fig.~\ref{fig:ssh_kitaev}(b), and the trivial phase ground state of $H_{\text{SSH}}(\delta = -1)$, shown in Fig.~\ref{fig:ssh_kitaev}(c).

One might expect that reaching a more entangled state—such as the ground state of Eq.~\eqref{eq:Hrandom}, corresponding to the red curve—would be the most complex. However, the results show that reaching the critical states at $\delta = 0$ (blue and brown curves, respectively) is actually more complex. Strikingly, the complexity of reaching the unentangled, fully occupied state shown in Fig.~\ref{fig:ssh_kitaev}(a) (pink curve) coincides with that of the random state (red curve).

Understanding the full implications of these results is beyond the scope of this work. Nevertheless, they support the idea that this definition of complexity does not necessarily follow the intuition that \emph{more entanglement implies more complexity}.

\section{Conclusions}
\label{sec:conclusions}

In this work, we have shown that the geometry of complexity obtained by minimizing the Riemmanian distance \eqref{eq:riemmanian_cost} in the covariance matrix approach is able to detect all the transitions that fidelity can, shedding some light into the controversy of whether complexity is a universal probe of quantum phase transitions or not. Furthermore, we have established that the complexity between ground states of very similar Hamiltonians reduces to the Fubini-Study distance. 

Also, we have provided non-fine-tuned examples where the entanglement entropy of a given subsystem provides more information than the complexity. One could argue that the comparison is not fair because the non-trivial information from the entanglement entropy arises when considering subsystems, but the complexity is expected to capture this. 

 This lack of more precise information is a direct consequence of the nature of the length functional $F[U(\tau),\boldsymbol{\alpha}(t)]$ \eqref{eq:riemmanian_cost}. An alternative definition of complexity, which is not subordinated to unitary complexity but is based on the Fubini-Study metric restricted to Gaussian states \cite{chapmanDefinitionComplexityQuantum2018} yields similar results.
 
 Indeed, penalizing all gates that do not generate Gaussian states is not enough. What is lacking in both approaches is the notion of spatial locality—locality in momentum space is not sufficient—and the consequent non-isotropic nature of the metric, which is manifested by imposing penalty factors $p_i$ that penalize directions (or gate choices) involving fermions that are not spatially close:
\begin{equation}
F[U(t), \boldsymbol{\alpha}(t)] \equiv \sqrt{\sum_i p_i \alpha_i^2}, \quad p_i \geq 0.
\label{eq:penalized_cost}
\end{equation}
 As recognized in \cite{jeffersonCircuitComplexityQuantum2017} and \cite{hacklCircuitComplexityFree2018}, this is a challenging problem.

In the study of the evolution of the entanglement entropy and the complexity in the SSH model \cite{aliPostquenchEvolutionComplexity2020}, the authors observed that the EE saturated much earlier than the complexity. However, they did not attribute this observation to the definition of complexity; instead, they considered it an intrinsic property of the model.

We restricted the analysis to the realm of fermionic Gaussian states, expecting that they would also be applicable to the bosonic case and even to interacting systems. The latter case is significantly more difficult to address in terms of the geometric complexity. In fact, even in the context of circuit complexity it is generally hard to rule out the existence of a shorter circuit. Nevertheless, there have been substantial advances in relating the hierarchy of randomness—namely, unitary $k$-designs—to circuit complexity in random quantum circuits \cite{brandaoLocalRandomQuantum2016,robertsChaosComplexityDesign2017,brandaoModelsQuantumComplexity2021}, including evidence for the conjectured linear growth of complexity. This behavior is likely a consequence of the local structure of the circuits, which are made up of nearest neighbors two-qubit unitaries.

We find the work in Ref. \cite{jayaramaRealspaceCircuitComplexity2023} particularly relevant, as it demonstrates that penalizing non-locality enhances the resolution of the phase diagram for the complexity of the XY model. It would be interesting to compare this notion of complexity with the information provided by entanglement entropy. Furthermore, it would be intriguing to explore whether the concept of subsystem complexity, defined as the complexity of a circuit where the target and reference states are reduced density matrices describing the same spatial subsystem \cite{digiulioSubsystemComplexityGlobal2021,digiulioSubsystemComplexityLocal2021}, can distinguish the various states considered in this work, even without explicitly imposing spatial locality in the functional. Finally, it would be worthwhile to investigate whether the recently observed insensitivity of (spread) complexity to the measurement rate \cite{sahuQuantumComplexityLocalization2024} arises from a similar absence of spatial locality in its definition.

\begin{acknowledgments}
It is a pleasure to thank Giuseppe de Tomasi, Pedro Ribeiro, Javier Molina Villaplana, Erik Tonni and Javier Rodríguez-Laguna for helpful discussions and contributions. The author acknowledges support from FCT-Portugal through Grant No. UID/CTM/04540/2020 and from DQUANT QuantEra II Programme through Grant No.101017733 https://doi.org/10.54499/QuantERA/0003/2021.
\end{acknowledgments}

\onecolumngrid
\appendix

\section{On covariance matrices and models}
\label{app:oncov_andmodels}

In this pedagogical appendix, we provide details on the derivation of the covariance matrices and the models discussed in the main text. Both the Kitaev and SSH models can be solved analytically by transforming into momentum space. For a comprehensive review of the SSH model, we refer to Ref. \cite{asbothSuSchriefferHeegerSSHModel2016}, and for the Kitaev chain—or equivalently, the quantum transverse Ising chain—we refer to Ref. \cite{mbengQuantumIsingChain2024}.  
Here, we review the case of a generic free-fermion Hamiltonian whose solvability is not feasible via a momentum-space transformation. We believe that this may be useful, at least from a numerical perspective. The complexity formalism discussed in Ref. \cite{hacklCircuitComplexityFree2018} is general and extends beyond quasi-translationally invariant systems.

Hence, we shall explain how to obtain the covariance matrix in the case where the system has particle number conservation (no pairing terms) and where it does not. We shall also provide expression in terms of Dirac and Majorana fermions.

\subsection{Correlation matrix of non-particle conserving systems}

Let us start by considering the Nambu spinor of $N$ fermions.
\begin{equation}
    \bm{\Psi}=\begin{pmatrix}
        c_1\\
        \vdots\\
        c_N\\
        c_1^\dagger\\
        \vdots\\
        c_N^\dagger
    \end{pmatrix}= \begin{pmatrix}
        \bm{c}\\
        (\bm{c^\dagger})^T
    \end{pmatrix},
    \label{eq:nambu}
\end{equation}
 Consider the most general free fermion Hamiltonian
 \begin{equation}
\hat{H}_{FF}=\sum_{j,j'}T_{j'j}c^\dagger_jc_{j'}+ \sum_{j,j'} F_{j'j}c^\dagger_jc^\dagger_{j'}+ F^*_{jj'}c_jc_{j'}+ \tr T,
\end{equation}
can be written compactly in terms of the single-body Hamiltonian $\mathbb{H}$
\begin{equation}
\hat{H}_{FF}=\frac{1}{2}\mathbf{\Psi}^\dagger \mathbb{H}\mathbf{\Psi}, \quad \mathbb{H}= \lp\begin{array}{c|c}
  \mathbf{T} & \mathbf{F} \\ 
  \hline
  -\mathbf{F}^* & -\mathbf{T}^*
 \end{array}\rp,
 \label{eq:app_generic_ham_dirac}
\end{equation}
where the hopping matrix is Hermitian $\mathbf{T}^\dagger=\mathbf{T}$ and the pairing matrix is skew-symmetric $\mathbf{F}^T=-\mathbf{F}$.

The two-point correlation function, the main quantity for this states takes the form
\begin{equation}
\ev{\bm{\Psi}\bm{\Psi}^\dagger}=\lp\begin{array}{c|c}
  \ev{\bm{c}(\bm{c}^\dagger)^T}& \ev{\bm{c}(\bm{c})^T} \\ 
  \hline
  \ev{(\bm{c}^\dagger)^T\bm{c}^\dagger}\ &\ev{(\bm{c}^\dagger)^T(\bm{c})^T}
 \end{array}\rp = \lp\begin{array}{c|c}
  \mathbb{1}-\bm{C}^T& \bm{G} \\ 
  \hline
  \bm{G}^\dagger &\bm{C}
 \end{array}\rp,
 \label{eq:app_corr}
\end{equation}
where $\bm{C}^\dagger=\bm{C}$ and $\bm{G}^\dagger=-\bm{G}^*$. 

Observe that $\mathbb{H}$ is invariant under a particle hole transformation, meaning that if $\lambda$ is an eigenvalue so it is $-\lambda$. Hence, there exist an unitary matrix $\bm{V}$ such that $\bm{V}\mathbb{H}\bm{V}^\dagger=\bm{D}$ with $\bm{D}=\text{diag}(\epsilon_1,\dots,\epsilon_N,-\epsilon_1,\dots,-\epsilon_N)$. Plugging it into \eqref{eq:app_generic_ham_dirac} we find
\begin{equation}
\hat{H}_{FF}=\frac{1}{2}\mathbf{\Psi}^\dagger \mathbb{H}\mathbf{\Psi}=\frac{1}{2}\mathbf{\Psi}^\dagger \bm{V}^\dagger\bm{D}\bm{V}\mathbf{\Psi} = \frac{1}{2}\mathbf{\Phi}^\dagger\bm{D}\mathbf{\Phi},
\label{eq:app_diag_ham_dirac}
\end{equation}
where we define a new Nambu spinor representing a new set of fermionic operators
\begin{equation}
    \bm{\Phi}=\begin{pmatrix}
        \bm{b}\\
        (\bm{b^\dagger})^T
    \end{pmatrix}.
\end{equation}
Crucially, the ground state of $\hat{H}_{FF}$ is the Fock space of these fermions $b_k\ket{GS}=0$ for $k=1,\dots,N$. Hence, the correlation matrix $ \ev{\bm{\Phi}\bm{\Phi}^T}$ takes the simple form 
 \begin{equation}
     \ev{\bm{\Phi}\bm{\Phi}^T}=\lp\begin{array}{c|c}
  \mathbb{1}& \mathbb{0} \\ 
  \hline
  \mathbb{0} &\mathbb{0}
 \end{array}\rp.
 \label{eq:app_corr_fock}
 \end{equation}
Hence, we obtain it expressed in the original fermions:
\begin{equation}
 \ev{\bm{\Psi}\bm{\Psi}^\dagger}=\bm{V}^\dagger\lp\begin{array}{c|c}
  \mathbb{1}& \mathbb{0} \\ 
  \hline
  \mathbb{0} &\mathbb{0}
 \end{array}\rp\bm{V}.
 \label{eq:app_corr_V}
\end{equation}

 It will be convenient to write also the correlation matrix in terms of Majorana fermions  $c_i=1/2(\alpha_1+i\beta_i)$ that satisfy the anticommutation relations $\{\alpha_i,\beta_j\}=0$, $\{\alpha_i,\alpha_j\}=\{\beta_i,\beta_j\}=2\delta_{ij}$:
\begin{equation}
\begin{split}
\mathbf{\Psi}=\mathbf{U}_M\bm{\xi}\equiv\frac{1}{2}\lp\begin{array}{c|c}
  \mathbf{\mathbb{1}}& i\mathbf{\mathbb{1}} \\ 
  \hline
  \mathbf{\mathbb{1}} &-i\mathbf{\mathbb{1}}
 \end{array}\rp\begin{pmatrix}\bm{\alpha}\\ \bm{\beta}\end{pmatrix},\\
   \mathbf{\xi}=\mathbf{U}^{-1}_M\bm{\Psi}=\lp\begin{array}{c|c}
  \mathbf{\mathbb{1}} & \mathbf{\mathbb{1}} \\ 
  \hline
  -i\mathbf{\mathbb{1}} &i\mathbf{\mathbb{1}}
 \end{array}\rp\begin{pmatrix}\bm{c}\\ (\bm{c}^\dagger)^T\end{pmatrix}
 \end{split}
 \label{eq:app_spinor_dirac_to_majo}
\end{equation}

Now using the above expression in \eqref{eq:app_corr} we find 
\begin{equation}
    \ev{\bm{\xi}\bm{\xi}^T}=\lp\begin{array}{c|c}
  \mathbb{1}& \mathbb{0} \\ 
  \hline
  \mathbb{0} &\mathbb{1}
 \end{array}\rp +i\bm{\Gamma},
\end{equation}
where we have extracted the symmetric part and therefore the covariance matrix is 
\begin{equation}
    i\bm{\Gamma}=\lp\begin{array}{c|c}
  \bm{C}-\bm{C}^T+\bm{G}-\bm{G}^*& i[\mathbb{1}-(\bm{C}+\bm{C}^T+\bm{G}+\bm{G}^*)] \\ 
  \hline
  -i[\mathbb{1}-(\bm{C}+\bm{C}^T-\bm{G}-\bm{G}^*)] & \bm{C}-\bm{C}^T+\bm{G}^*-\bm{G}
 \end{array}\rp.
 \label{eq:app_cov_CandG}
\end{equation}
 Notice that since $\bm{C}-\bm{C}^T$ and $\bm{G}-\bm{G}^*$ are purely imaginary and antisymmetric, the covariance matrix inherits this property.

 Using \eqref{eq:app_spinor_dirac_to_majo} we can also define a new set of Majorana fermions $\bm{\chi}$ for which the covariance matrix is 
  \begin{equation}
     \bm{\Gamma}_\text{Fock}=\lp\begin{array}{c|c}
  \mathbb{0}& \mathbb{1}\\ 
  \hline
  -\mathbb{1}& \mathbb{0}
 \end{array}\rp \ ,
 \label{eq:app_cov_fock}
 \end{equation}
 
 which can be seen by making $\bm{C}=\bm{G}=0$ in \eqref{eq:app_corr}
 or by rewritting in the Majorana basis using \eqref{eq:app_spinor_dirac_to_majo}  the Fock correlation matrix \eqref{eq:app_corr_fock}.
 
 A summary of the formalism in both Dirac and Majorana fermions can be seen below
\begin{figure}[h!]
    \centering
    \includegraphics[width=0.65\linewidth]{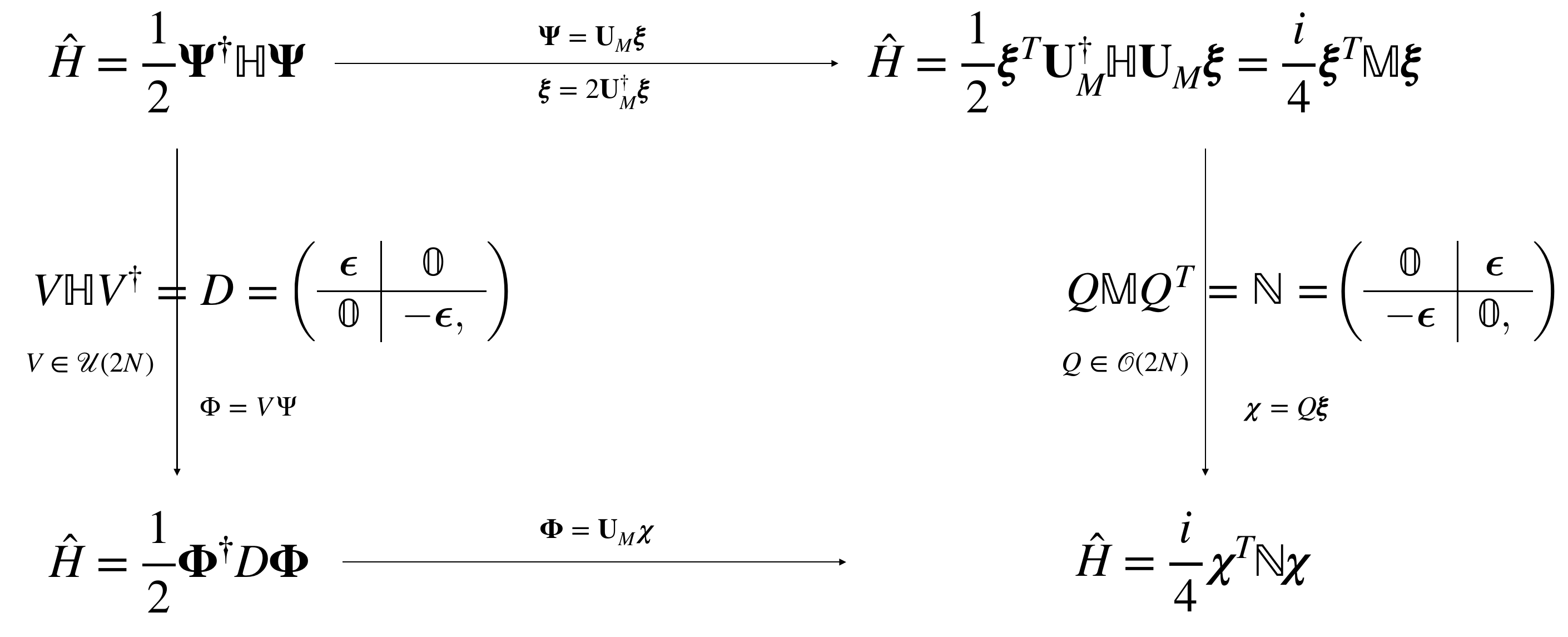}
    \end{figure}

Finally, observe that a generic Hamiltonian in Majorana basis is given by
\begin{equation}
    \hat{H}=\frac{i}{4}\bm{\xi}^T\mathbb{M}\bm{\xi}, \quad \mathbb{M}^T=-\mathbb{M}.
\end{equation}

There exists a unitary $\bm{W}$ such that $\bm{W}i\mathbb{M}\bm{W}^\dagger=\bm{D}$. Since $\bm{U}^\dagger_M\mathbb{H}\bm{U}_M=\frac{i}{2}\mathbb{M}$, and we established in \eqref{eq:app_diag_ham_dirac} that $\bm{V}\mathbb{H}\bm{V}^\dagger=\bm{D}$ we have that $\bm{V}=\bm{W}\bm{U}^\dagger_M$.  Using \eqref{eq:app_corr_V} we have that
\begin{equation}
   \ev{\bm{\Psi}\bm{\Psi}^\dagger}=\bm{V}^\dagger\lp\begin{array}{c|c}
  \mathbb{1}& \mathbb{0} \\ 
  \hline
  \mathbb{0} &\mathbb{0}
 \end{array}\rp\bm{V}=\bm{U}_M\ev{\bm{\xi}\bm{\xi}^T}\bm{U}_M^\dagger. 
\end{equation}
Therefore the covariance matrix is
\begin{equation}
    \text{Im}\lp\bm{U}^{-1}_{M}\bm{V}^\dagger\lp\begin{array}{c|c}
  \mathbb{1}& \mathbb{0} \\ 
  \hline
  \mathbb{0} &\mathbb{0}
 \end{array}\rp\bm{V}(\bm{U}^\dagger_M)^{-1}\rp =\text{Im}\lp\bm{W}^\dagger\lp\begin{array}{c|c}
  \mathbb{1}& \mathbb{0} \\ 
  \hline
  \mathbb{0} &\mathbb{0}
 \end{array}\rp\bm{W}\rp
\end{equation}
 \begin{figure}

    \centering
    \includegraphics[width=1\linewidth]{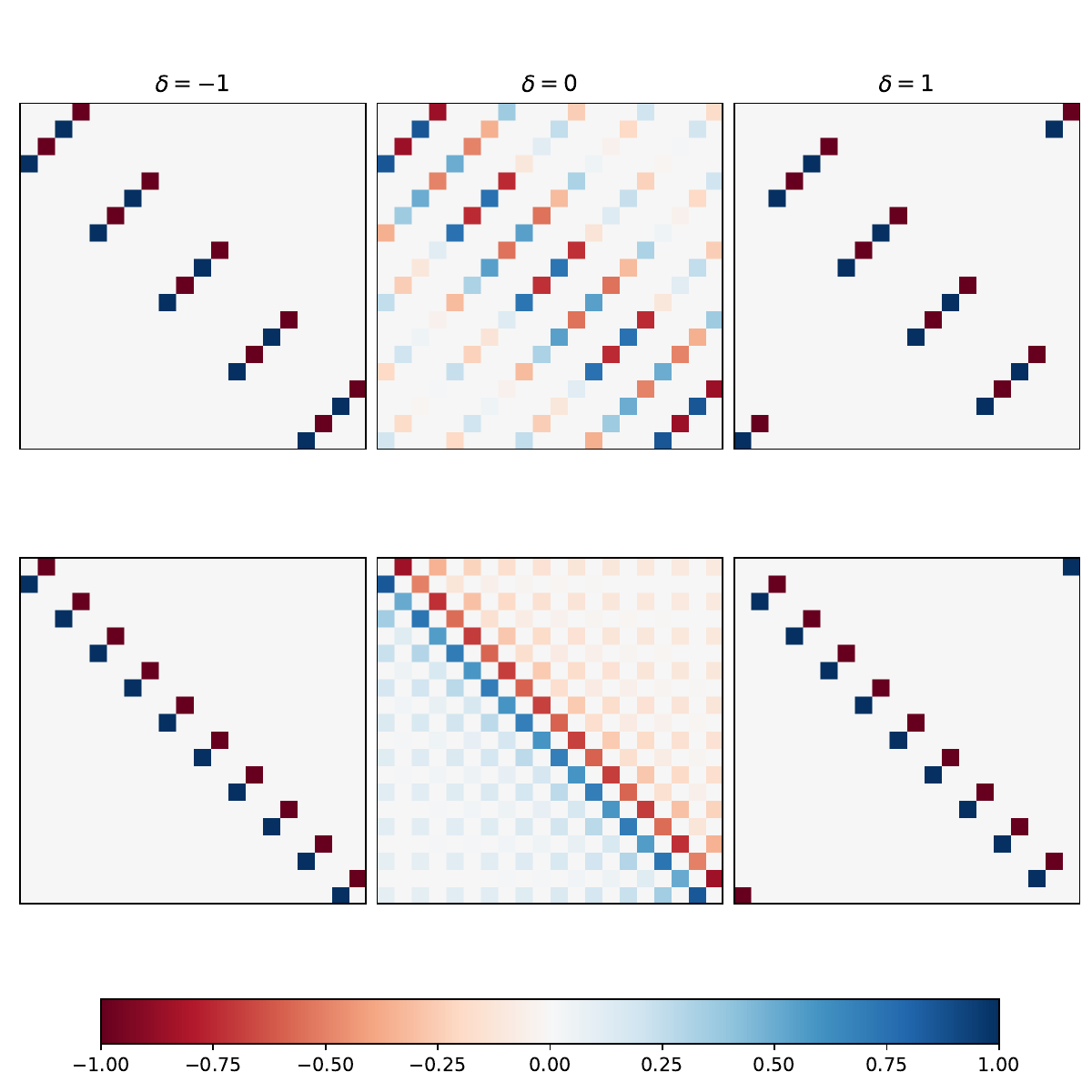}
    \caption{Covariance matrices of the SSH (upper row) and Kitaev (lowe row) for representative values of  $\delta$.  Notice that we have expressed it in terms of a Majorana spinor that makes spatial locality explicit $(\alpha_1,\beta_1,\dots\alpha_N,\beta_N)$.}
    \label{fig:covs_sshkitaev}
\end{figure}
\subsection{Correlation matrix of particle conserving systems}
Let us start with the simplest case of a free fermion Hamiltonian given by:
\begin{equation}
    H_{pc}=\sum_{jj'}T_{jj'}c_j^\dagger c_{j}+T_{jj'}\delta_{jj'} = \bm{c}^\dagger\bm{T}\bm{c}+\tr\bm{T}, \quad \bm{T}^\dagger=T
\end{equation}
There exists an unitary $\bm{u}$ that diagonalizes the hopping matrix $\bm{u}^\dagger\bm{T}\bm{u}=\bm{\epsilon}\equiv\text{diag}(\epsilon_1,\dots,\epsilon_N)$. The Hamiltonian is diagonal when written in terms of the fermions $\bm{b}=\bm{u}^\dagger\bm{c}$, or $$b_k=\sum_{j}u^*_{jk}c_j.$$
Crucially, both families of fermions share the same Fock vacuum.  The ground state $\ket{GS}=\prod_{\Omega_S}b^\dagger_k\ket{0}$ with $\Omega=\{k \ | \ \epsilon_k<0 \}$. The cardinal of the $\Omega$ is $n_f$ and, in systems with particle hole or sublattice symmetry, it is always $n_f=N/2$ (half filling). If there are zero modes, $n_f$ and consequenlty the parity can differ but not the total energy of the ground state.
The two-point correlation matrix is 
\begin{equation}
    \mel{GS}{c^\dagger_ic_j}{GS}=\sum_{kk'}U^*_{ik}U_{jk'}\mel{GS}{b^\dagger_k b_{k'}}{GS}=\sum_{k=1}^{n_F}U^*_{ik}U_{jk}
\end{equation}
Or in a matrix form
\begin{equation}
    \bm{C}\equiv\ev{(\bm{c}^\dagger)^T(\bm{c})^T} =(P_{nf}\bm{u}^T)^\dagger P_{nf}\bm{u}^T= \bm{u}^*P_{nf}\bm{u}^T, \quad P_{nf}=\lp\begin{array}{c|c}
  \mathbb{1}_{nf}& \mathbb{0}\\ 
  \hline
  \mathbb{0}& \mathbb{0}
 \end{array}\rp
 \label{eq:corr_pc}
\end{equation}

Hence,  we can obtain the covariance matrix (in Majorana basis) for this kind of systems by particularizing \eqref{eq:app_cov_CandG}  with we find $\bm{G}=0$ and $\bm{C}$ computed from \eqref{eq:corr_pc}

\begin{equation}
    \bm{\Gamma}_{pc}=\lp\begin{array}{c|c}
  \mathbb{0}& [\mathbb{1}-2\bm{C}] \\ 
  \hline
  -[\mathbb{1}-2\bm{C}] & \mathbb{0}
 \end{array}\rp.
\end{equation}
\subsection{Covariance matrix of the SSH and Kitaev models}

Finally, it is worth to apply the above to the specific examples of the SSH and Kitaev models. Since we have chosen to work in the Majorana basis, let us start by writing the SSH chain in terms of these fermions.
It is easy to see that given
\begin{equation}
    H_{\text{SSH}}=-\sum_{m=1}^{N-1}(1+(-1)^m\delta)\left(c^\dagger_mc_{m+1} \ + \ \text{h.c}\right),
    \label{eq:app_ssh_ham}
\end{equation}
we can rewrite it as
\begin{equation}
    H_{\text{SSH}}=-\frac{i}{4}\sum_{m=1}^{N-1}(1+(-1)^m\delta)(\alpha_m\beta_{m+1}-\beta_m\alpha_{m+1}).
\end{equation}
The Kitaev chain is
\begin{equation}
    H_\text{K}=\frac{i}{4}\sum_{m=1}^{N-1}(1-\delta)\alpha_m\beta_m+ (1+\delta)\beta_m\alpha_{m+1} +(1-\delta)\alpha_N\beta_N,
\end{equation}

In Fig. \ref{fig:covs_sshkitaev} we plot the covariance matrices for the clean (generalized) bond states cases $\delta=\pm1$ and the critical case $\delta=0$.  Notice that we have expressed in terms of a spinor that makes spatial locality explicit $$(\alpha_1,\dots\alpha_N,\beta_1,\dots,\beta_N)\to (\alpha_1,\beta_1,\dots\alpha_N,\beta_N).$$
Notice how the non-trivial state $\delta=1$ displays a non-local state.
\section{Comparing fidelity and complexity}
\label{app:complexity_and_fidelity}
It is straightforward to see that
\begin{equation}
    \dot{\mathcal{C}}(\delta)=-\frac{1}{2}\dfrac{\sum_{j=1}^{2N} \theta_j(\delta)\dot{\theta_j}(\delta)}{\mathcal{C}(\delta)},
\end{equation}
implying that if $\delta_c$ is an extreme, then $\dot{\theta_j}(\delta_c)=0$ for all $j$. Concerning the second derivative,
\begin{equation}
    \ddot{\mathcal{C}}(\delta)=-\frac{1}{\mathcal{C}(\delta)}\left(\dot{\mathcal{C}}^2(\delta)+ \sum_{j=1}^{2N}\ddot{\theta}(\delta)\theta_j(\delta)\right),
\end{equation}
that diverges in the critical point if at least one $\ddot{\theta_j}(\delta_c)\to\infty$. 

The derivative of the fidelity is
\begin{equation}
\begin{split}
    \dot{\mathcal{F}}(\delta)&=\frac{1}{\mathcal{F}(\delta)}\sum_{j=1}^{2N}\sin\frac{\theta_j(\delta)}{2}\dot{\theta(\delta})\prod_{i\neq j}\cos\frac{\theta_i(\delta)}{2}\\
    &=\mathcal{F}(\delta)\sum_{j=1}^{2N}\tan\frac{\theta_j(\delta)}{2}\dot{\theta(\delta}).
\end{split}
\end{equation}
Clearly, if $\dot{\theta_j}(\delta_c)=0$ for all $j$ the fidelity has also an extrema at $\delta_c$. Concerning the second derivative,  
\begin{equation}
\begin{split}
    \ddot{\mathcal{F}}(\delta)&=\dot{\mathcal{F}}(\delta)\sum_{j=1}^{2N}\tan\frac{\theta_j(\delta)}{2}\dot{\theta(\delta})\\
    &\quad \quad +\mathcal{F}(\delta)\sum_{j=1}^{2N}2\sec^2\frac{\theta_j(\delta)}{2}\dot{\theta}^2(\delta)+\tan\frac{\theta_j(\delta)}{2}\ddot{\theta}(\delta),
\end{split}
\end{equation}
which also diverges if at least one $\ddot{\theta_j}(\delta_c)\to\infty$. Hence, we find that all phase transitions that complexity may detect are also probed with fidelity.
\bibliography{all_zotero}

\end{document}